\documentclass[a4paper,number,5p,twocolumn]{elsarticle}
\usepackage{graphicx}
\usepackage{natbib}
\usepackage{a4wide}
\usepackage[centertags]{amsmath}
\usepackage{amssymb}
\usepackage{latexsym}
\usepackage{subfigure}
\usepackage[]{hyperref}
\begin{document}
%
\author[UR]{C. Fortmann}
\ead{carsten.fortmann@uni-rostock.de}
\author[UR]{R. Thiele}
\author[DESY]{R. R. F\"austlin}
\author[UR]{Th. Bornath}
\author[UR]{B. Holst}
\author[UR]{W.-D. Kraeft}
\author[UR]{V. Schwarz}
\author[DESY]{S. Toleikis}
\author[DESY]{Th. Tschentscher}
\author[UR]{R. Redmer}
%
\address[UR]{Institut f\"ur Physik, Universit\"at Rostock, 18051
Rostock, Germany}
\address[DESY]{Deutsches Elektronensynchrotron DESY, Notkestrasse 85, 22607 Hamburg, Germany}
%
\title{Thomson scattering in dense plasmas with density and temperature gradients}
%
\begin{abstract}
  Collective X-ray Thomson scattering has become a versatile tool for the diagnostics of dense plasmas. 
  Assuming
  homogeneous density and temperature throughout the target sample, 
  these parameters can be determined directly from the plasmon dispersion and the
  ratio of plasmon amplitudes via detailed balance. 
  In inhomogeneous media, the scattering signal is an average of the density and temperature
  dependent scattering cross-section
  weighted with the density and temperature profiles. 
  We analyse Thomson scattering spectra in the XUV range from
  near solid density hydrogen targets generated by free electron laser radiation.
  The influence of plasma inhomogeneities on the scattering spectrum is 
  investigated by comparing density and temperature
  averaged scattering signals to calculations assuming homogeneous targets. 
  We found discrepancies larger than 10\% between the mean electron density and the
  effective density
  as well as between the mean temperature and the effective
  temperature.
\end{abstract}
%
\maketitle

\section{Introduction\label{sec:intro}}
Recently, X-ray Thomson scattering has been demonstrated to be a reliable and versatile tool for the diagnostics of dense plasmas
\cite{GlenzerRedmer_RMP_2009}. Examples for applications cover a large range from laser produced plasmas at 
temperatures of several keV and densities
between solid density and a few percent of solid density (hot dense plasmas)\cite{Glenzer_PhysRevLett.82.97_1999,Gregori_JQSRT} via so-called warm dense matter (WDM) 
\cite{Lee_LPB20_527_2002,Lee_JOSA20_770_2003} with temperatures of several eV and densities close
to solid density \cite{glen:prl03, Glenzer:PRL98_065002_2007}, up to compressed matter at temperatures between 0.1 eV and
several 10 eV \cite{Glenzer_JPhysConf8_112_032071,Lee_PRL_submitted}.

The determination of the plasma properties, such as the plasma density, the plasma temperature, and the plasma composition, requires the knowledge
of the density and temperature dependent scattering cross-section. As will be outlined in Sec.~\ref{sec:theory}, 
the cross-section for Thomson scattering is usually expressed via the
dynamical structure factor (DSF) $S_\mathrm{ee}(\mathbf{k},\omega)$  \cite{Chihara_JPhysCondMat12_231_2000},
which is a function of the scattering wavevector
$\mathbf{k}$ and the frequency shift $\omega$, convoluted with the density and temperature distribution of the target.
In this way, plasma inhomogeneities can be taken into account in the simulation of the scattering spectrum, which is the main purpose of this
paper.
%
Such analysis has been performed earlier by Baldis et al. \cite{bald:rev.sci.inst02} for the case of 
VUV photons interacting with solid aluminum targets. Here, the scenario of XUV photons delivered by the free electron laser facility at
DESY-Hamburg (FLASH) \cite{Ackermann_NaturePhotonics1_336_2007} scattering on liquid hydrogen at near solid density (mass density
$\rho=0.088\,\mathrm{g/cm^{3}}$) as proposed by
H\"oll et al. \cite{Hoell:HEDP3_2007}, will be considered. 

%
\section{Theory of Thomson Scattering and the dynamic structure factor\label{sec:theory}}
The scattered power per solid angle $d\Omega=\sin\vartheta\,d\vartheta\,d\varphi$ and unit frequency interval $d\omega$ is given by the expression \cite{bald:rev.sci.inst02}
\begin{multline}
  \frac{d^2P_\mathrm{sc}}{d\Omega\,d\omega} = \frac{\sigma_\mathrm{T}}{A_\mathrm{rad}}
  \frac{k_\mathrm{f}}{k_\mathrm{i}}
  \int\limits_{-\infty}^{\infty}\!\!\frac{d\omega'}{2\pi}
  G^{}_{\Delta\omega}(\omega-\omega')\,
  \\
  \times\int\limits_{\mathcal V_\mathrm{rad}}\!\! d^3\mathbf{r}\,I(\mathbf{r})\,
  S_\mathrm{ee}(\mathbf{k},\omega';n_\mathrm{e}(\mathbf{r}),T(\mathbf{r}))\,n_\mathrm{e}(\mathbf{r})~.
  \label{eqn:d2PdwdW}
\end{multline}
$\sigma_\mathrm{T}=6.65\times 10^{-24}\,\mathrm{cm^{2}}$ is the Thomson cross-section, $\mathbf{k_\mathrm{i}}$ and $\mathbf{k}_\mathrm{f}$ 
are the initial and the final photon wavevector, respectively. By $\mathbf{k}=\mathbf{k}_\mathrm{i}-\mathbf{k}_\mathrm{f}$,
we denote the scattering wavevector. From the conservation of momentum, we obtain $k=4 \pi\sin(\vartheta/2)/\lambda_0$, with $\lambda_0$ being
the probe wavelength. $I(\mathbf{r})$ is the $\mathbf{r}$ dependent power density of
incoming photons, while $A_\mathrm{rad}$ is the radiated surface of the target. 
%
%
Via the convolution of the DSF with an appropriately chosen instrumental
function $G^{}_{\Delta\omega}(\omega)$, the finite resolution of the spectrometer as well as the probe's bandwidth is taken into account.
Here, we use the normalized Gaussian distribution
\begin{equation}
  G^{}_{\Delta\omega}(\omega)=\frac{1}{\sqrt{2\pi}\sigma}\exp\left( -\frac{\omega^2}{2\sigma^2} \right)~,
  \label{eqn:instrumental_function}
\end{equation}
$\Delta\omega= 2\sqrt{2\ln 2}\,\sigma$ is the full width at half maximum (FWHM) of the instrumental function.

Finally, the plasma inhomogeneities are taken into account by averaging
the DSF with the electron density profile $n_\mathrm{e}(\mathbf{r})$ and the temperature profile $T(\mathbf{r})$. 

It is convenient to separate the DSF into contributions from free-free, free-bound and bound-bound correlations as proposed by Chihara
\cite{Chihara_JPhysCondMat12_231_2000},
\begin{multline}
  S_\mathrm{ee}(\mathbf{k},\omega)=
  \left|f_\mathrm{i}(k)+q(k)\right|^2\,S_\mathrm{ii}(\mathbf{k},\omega)\\
  +Z_\mathrm{c}\int_{-\infty}^{\infty}d\omega'\,S_\mathrm{c}(\mathbf{k},\omega)S_\mathrm{s}(\mathbf{k},\omega-\omega')\\
  +Z_\mathrm{f}S_\mathrm{ee}^{0}(\mathbf{k},\omega)~.
  \label{eqn:chihara_decomp}
\end{multline}
The first part gives the correlation of electrons that are weakly and tightly 
bound to the ions and follow the ion's
movement adiabatically. The amplitude is
determined by the atomic form factor $f_\mathrm{i}(k)$, 
i.e. the charge distribution of the electrons in the valence shell orbitals and
the 
screening cloud $q(k)$ which gives the distribution of quasi-free electrons screening the ion's charge \citep{Gregori:HEDP3_2007}. 

The second term contains the contribution of core electrons, $S_\mathrm{c}(\mathbf{k},\omega)$ and
describes Raman type transitions of inner shell electrons to the continuum,
modulated by the ion's movement which is contained in $S_\mathrm{s}(\mathbf{k},\omega)$ \citep{SahooRiley_PRE77_046402_2008}.

Finally, $S_\mathrm{ee}^{0}(\mathbf{k},\omega)$ is the free electron contribution.
It determines the behaviour of the total electron structure factor at
frequencies close to the plasma frequency. Since we focus on the application of Thomson scattering to diagnostics of dense
plasmas via analysis of the plasmon feature, the third term in Eq.~(\ref{eqn:chihara_decomp}) will be discussed in more detail in the following.
This contribution is calculated in the Born-Mermin approximation \cite{Redmer_IEEEPS33_2005}, thereby including
collisions among electrons and ions in second order Born approximation. Higher order terms can be treated via the t-matrix and inclusion
of dynamical screening, see Ref. \cite{Thiele_JPhysA_39_2006}. The influence of collisions on the collective scattering spectrum has been
thoroughly discussed by some of the authors in Ref.~\cite{Fortmann_LPB_2009}. In this paper, we will concentrate on the impact of the density
and temperature profiles on the scattering signal.

%
\section{Temperature and density profiles\label{sec:profiles}}
\begin{figure}[ht]
  \begin{center}
    \includegraphics[width=.49\textwidth,angle=0,clip]{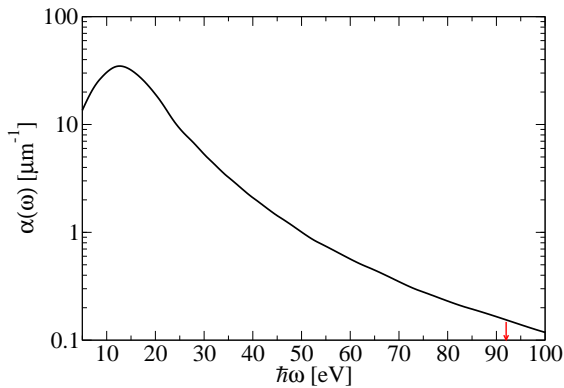}
  \end{center}
  \caption{(Color online) Ab-initio simulation of the inverse absorption length for cryogenic hydrogen at solid density as a function of the photon energy.
  The red arrow indicates the  FLASH energy (92 eV).}
  \label{fig:absorption_H_FLASH}
\end{figure}

We consider the interaction of cryogenic ($T\simeq 20\,\mathrm{K}$) liquid hydrogen droplets of about $30\,\mathrm{\mu m}$ diameter
with XUV free electron laser radiation delivered by FLASH at 
$13.5\,\mathrm{nm}$ wavelength with a pulse length of $30\,\mathrm{fs}$ and $50\,\mathrm{\mu J}$ pulse energy. 
In order to infer the absorption characteristics of the target, we calculate 
the inverse absorption length $\alpha(\omega)$ of cryogenic hydrogen ($T=20\,\mathrm{K}$) via ab-initio quantum molecular dynamic simulations using the Kubo-Greenwood formula \cite{Holst_PRB77_184201_2008}.
The inverse absorption length as a function of the photon energy is shown in
Fig.~\ref{fig:absorption_H_FLASH}.
At the FLASH energy ($92\,\mathrm{eV}$, indicated by the red arrow) 
the absorption length is $15\,\mathrm{\mu m}$, 
in excellent agreement with tabulated x-ray absorption data 
\cite{Henke_AtDataNuclDataTables54.181_1993}, which is in the same
order of magnitude as the diameter of the hydrogen droplet.
Thus, FLASH photons can penetrate deeply into the target.

The interaction of FLASH photons with the hydrogen target
was simulated using the radiation-hydrodynamic simulation code HELIOS \cite{MacFarlane_JQSRT99_381_2006}. HELIOS 
features a Lagrangian reference frame, separate ion and electron temperatures,
and flux-limited Spitzer thermal conductivity.
It allows for deposition of laser energy via inverse bremsstrahlung as 
well as bound-bound and bound-free transitions, using a SESAME-like 
equation of state.

\begin{figure}[ht]
  \begin{center}
    \subfigure[Density]{%
    \includegraphics[width=.49\textwidth,angle=0,clip]{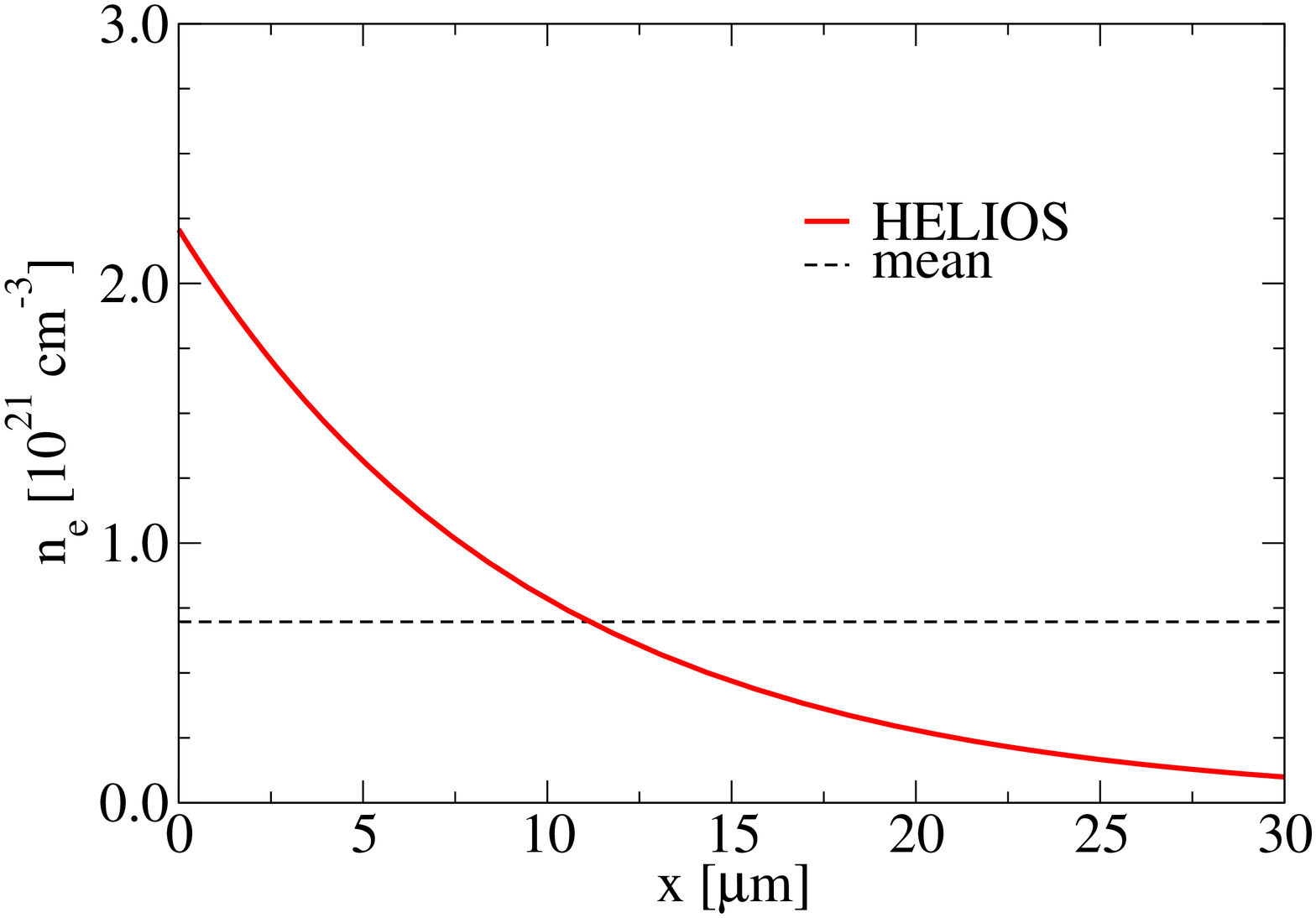}
    \label{subfig:profile_n}
    }
    \subfigure[Temperature]{%
    \includegraphics[width=.49\textwidth,angle=0,clip]{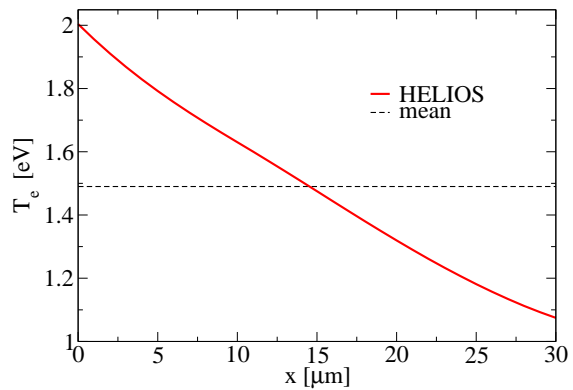}
    \label{subfig:profile_T}
    }
  \end{center}
  \caption{(Color online) Density (a) and temperature (b) profile from HELIOS simulation of liquid hydrogen droplet irradiated by XUV-FEL pulses,
  $\lambda_0=13.5\,\mathrm{nm}$, pulse length $30\,\mathrm{fs}$, pulse energy $50\,\mathrm{\mu J}$ (bold red curves). The dashed curves
  indicate the corresponding mean density and temperature.} 
  \label{fig:profile}
\end{figure}
Figure~\ref{fig:profile} shows the free electron density profile (a) and temperature profile (b) (red solid curves). 
The density profile decreases exponentially from $n_\mathrm{e}=2.2\times 10^{22}\,\mathrm{cm^{-3}}$ at the irradiated surface ($x=0$) to $n_\mathrm{e}=1.0\times
10^{20}\,\mathrm{cm^{-3}}$ at the rear surface ($x=3.0\,\mathrm{\mu m}$).
The 1/e-decay length of the exponential profile is roughly 11$\,\mu$m, 
which is consistent with the absorption length
obtained from the QMD simulation, c.f. Fig.~\ref{fig:absorption_H_FLASH}, i.e. the
absorption is mainly due to
bound-free transitions.

The plasma temperature decreases from $T=2\,\mathrm{eV}$ to $T=1.1\,\mathrm{eV}$.
For comparison, the dashed lines indicate the mean electron density $\bar
n_\mathrm{e}=7.0\times 10^{21}\,\mathrm{cm^{-3}}$ and mean temperature $\bar T=1.5\,\mathrm{eV}$,
respectively.
%
\section{Results\label{sec:results}}
\begin{figure}[ht]
  \begin{center}
    \includegraphics[width=.49\textwidth,angle=0,clip]{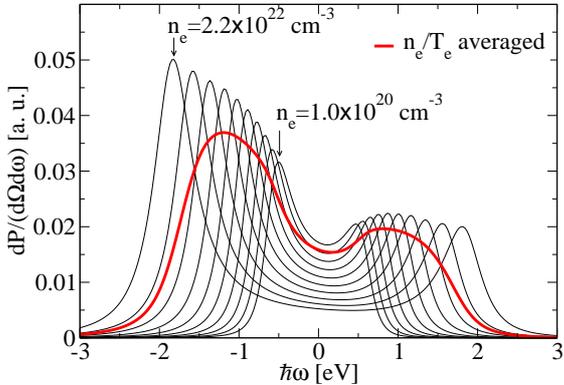}
  \end{center}
  \caption{(Color online) Thin curves: Scattered intensity spectra at constant density and temperature corresponding to the HELIOS profiles (Fig.~\ref{fig:profile}).
  Bold red curve: Density and temperature averaged scattering spectrum. Scattering angle $\vartheta=90^\circ$.
  }
  \label{fig:FlashFlash} 
\end{figure}

\begin{figure}[ht]
  \begin{center}
    \includegraphics[width=.49\textwidth,angle=0,clip]{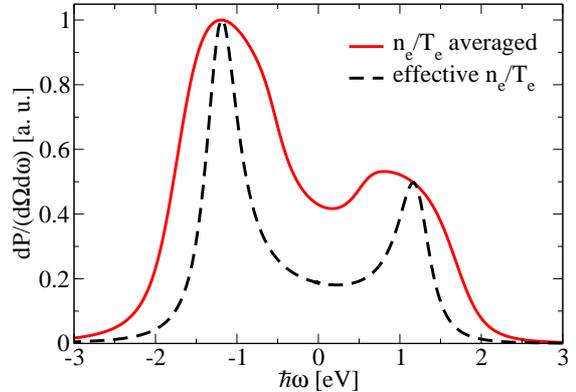}
  \end{center}
  \caption{(Color) Density and temperature averaged scattering spectrum (bold red curve) compared to
  a calculation using the effective density and temperature (dashed curve)
  obtained from the plasmon peak positions and detailed balance analysis from the profile averaged spectrum. Scattering angle $\vartheta=90^\circ$.}
  \label{fig:Skw_av-mean-eff}
\end{figure}

In Fig.~\ref{fig:FlashFlash}, 
we show calculations of the scattered intensity as a function of the photon energy shift with respect to the FLASH photon energy, 
assuming FLASH photons scattering from partially ionized hydrogen under an angle of $\vartheta=90^\circ$.
The thin curves show calculations of the scattering spectrum using a single density and
temperature combination corresponding to
a particular position $x$ in the HELIOS profiles, see Fig.~\ref{fig:profile}. Scattering from plasma regions close to the front surface (high density, high
temperature) results in a comparatively large plasmon resonance energy, while scattering from the rear surface (low density, low temperature) gives rise to small
plasmon energies, as indicated by the arrows and the corresponding electron densities. Finally, the bold red curve 
gives the density and temperature averaged scattered intensity, c.f. Eq.~(\ref{eqn:d2PdwdW}). 
Note, that since the DSF is weighted with the local density $n_\mathrm{e}(\mathbf{r})$, i.e. with 
the number of scattering electrons in the unit volume $d^3\mathbf{r}$, as well as with the
number of photons in that volume, regions close to the irradiated surface 
with increased density contribute more
strongly to the spectrum than the dilute parts located towards the rear surface.
This results in a scattering spectrum that is significantly more broadened than
the original spectra taken at constant density and temperature.

We compare the averaged spectrum to a calculation for the DSF
using homogeneous density and temperature distributions. The result is shown in 
Fig.~\ref{fig:Skw_av-mean-eff}. 
The bold red curve is again the DSF averaged with the
HELIOS profiles as in  Fig.~\ref{fig:FlashFlash}, the black dashed curve gives the DSF using the effective electron density
$n_\mathrm{eff}=8.3\times10^{20}\mathrm{cm^{-3}}$ and effective temperature $T_\mathrm{eff}=1.7\,\mathrm{eV}$, 
which are inferred via the peak position of the
plasmon resonances and the detailed balance analysis of the profile-averaged spectra (red curve). 
Both, effective density and effective temperature deviate significantly (i.e. more than 10\%) from the corresponding 
mean values $\bar n_\mathrm{e} = 7.0\times
10^{20}\,\mathrm{cm^{-3}}$ and
$\bar T=1.5\,\mathrm{eV}$.
Also note that in different setups, e.g. excitation of the plasma via intense optical laser
radiation, stronger density and temperature gradients are observed
\cite{Cao_LPB25.239_2007}, and even larger deviations between the effective values and
the mean values for $n$ and $T$ may occur.
 A further interesting observation is that the peaks of the averaged
 spectrum (Fig.~\ref{fig:Skw_av-mean-eff}) 
 are not symmetrically located with respect to 
 the incident photon energy as would be the case for a homogeneous 
 temperature and density distribution in the target. 
\section{Conclusions\label{sec:conclusions}}
We simulated Thomson scattering spectra including density and temperature gradients in the target. By comparing these spectra to calculations that assume
homogeneous density and temperature distributions, we found significant differences between these methods. In particular, the free electron density that is inferred from
the simulated spectrum via the plasmon dispersion, as well as the effective temperature,
inferred via detailed balance analysis of the simulated spectrum may not be interpreted as
the mean values of these observables. In the scenario studied here (FLASH photons at 13.5 nm
wavelenght scattering on near solid density hydrogen heated by the same photons),
both $n_\mathrm{eff}$ and $T_\mathrm{eff}$ differ by more than 10\% from the corresponding mean value.

Besides being important for the determination of density and temperature as demonstrated in this paper,
deconvolution of the scattering spectra and the profiles is also crucial if
one is interested in the broadening of the plasmon satellites due to collisions and Landau damping \cite{Fortmann_LPB_2009}.
Collisional broadening of plasmons gives valuable information about the electrical and
thermal conductivities in the plasma and is thus an important feature of the
scattering spectrum. However, neglecting the homogeneous broadening due to
the density and temperature gradients may lead to significant underestimation of the conductivity.
Thus, 
gaining information about the density and temperature profile of the target, e.g. via radiation hydrodynamic simulations,
is indispensable for the application of Thomson scattering spectra to plasma diagnostics. 

\section*{Acknowledgments}%
  This work was partly supported by the Deutsche Forschungsgemeinschaft (DFG) under Grant SFB 652 ``Strong correlations and collective phenomena in radiation
  fields: Coulomb systems, Cluster, and Particles'', by the
  DFG-Graduiertenkolleg 1355 (RRF), and
  the German Federal Ministry of Education and Research (BMBF) under Grant FSP 301-FLASH, project No. 05 KS7HRA.


\end{document}